\documentclass[prb,aps,epsfig,twocolumn,showpacs]{revtex4-1}
\usepackage{color}
\usepackage{epsfig,amssymb,amsmath,latexsym}
\usepackage{subfigure}
\usepackage{wrapfig}
\usepackage{float}
\usepackage{color}
\usepackage{bm,bbm,dsfont,ulem,nicefrac}
\usepackage[colorlinks=true,linktoc=page,linkcolor=magenta,citecolor=magenta]{hyperref}

\newcommand\bea{\begin{eqnarray}}
\newcommand\eea{\end{eqnarray}}
\newcommand\beq{\begin{equation}}  
\newcommand\eeq{\end{equation}}

\newcommand{\bib}{\bibitem}

\begin{document}
\title{\bf{ Quantum charge pumping through Majorana bound states }}
\author{ Krashna Mohan Tripathi$^{1}$, Sumathi Rao$^{1}$   and   Sourin Das$^{2,}$$^{3}$}
\affiliation{$^1$Harish-Chandra Research Institute, HBNI, Chhatnag Road, 
Jhusi, Allahabad 211 019, India. \\
$^2$Department of Physical Sciences, Indian Institute of Science Education and Research, \\ Kolkata, West Bengal,  741252, India.\\
$^3$Department of Physics and Astrophysics, University of Delhi, Delhi - 110 007, India.}

\begin{abstract}

We study adiabatic charge pumping through a Majorana bound state tunnel coupled to multiple normal leads. We show that for most of the parameters such a pump does not lead to any net pumped charge between the various leads unless a multiply connected geometry is implemented. We introduce an Aharonov-Bohm ring geometry at the junction to implement such a multiply connected geometry. We further show that the Fourier transform of the pumped charge with respect to flux inserted through the ring shows a clear distinction between the case of an Andreev bound state and the Majorana bound state. Hence such a Fourier analysis can serve as a diagnostic for the detection of Majorana bound states in the proposed geometry.

 \end{abstract}
 
\maketitle

\section{Introduction}

One of the first steps that is required for the application of Majorana modes\cite{Alicea,Beenakker} in 
quantum computation is  its  unambiguous identification. This has proved difficult
in experiments,\cite{Mourik,Majoranaexpts} since the usual diagnostic of Majorana bound states (MBS),   a zero bias peak in the conductance,  
can have many other origins besides signaling the presence of  a Majorana mode.  This fact has led to considerable work\cite{Recent} in recent
years, encompassing study of various toy models\cite{Toys} and promising physical systems\cite{Fukane,Sau} and their electrical transport signatures\cite{Noise}.

However, there has been no definitive experimental confirmation so far, which has proved the existence
of Majorana modes in any system and hence, it is still of interest to look for different ways to confirm the existence of Majorana modes.  
In this context, it is worth exploring the question of charge pumping through a Majorana mode 
and examining whether there are unique signals which can identify it and differentiate it
from pumping through other resonant levels or Andreev bound states. Charge pumping or the phenomenon of obtaining current in the absence of bias by local variations of
parameters of the quantum system, has been studied in many contexts, beginning with Thouless\cite{Thouless} who
considered the effect of a travelling periodic potential that could drag the electrons along. The analysis performed by Thouless was in the spirit of 
closed quantum system. 

Later the idea of pumping was extended to open quantum system in the Refs.~\onlinecite{BTP,Brouwer} where the pumping of charge is induced between 
different electron reservoirs by periodically varying independent parameters of scattering matrix that describes the scattering of electrons between the different 
electron reservoirs. When the variation of the parameters is much slower than the transport time, then the pumping is
adiabatic and the Brouwer formula\cite{Brouwer} can be applied. Adiabatic charge pumping has
attracted a great deal of interest in the last several years, and different aspects of it have been studied in great detail\cite{Aleiner, Zhou, Wei, Levinson, Renzoni, 
Blaauboer2001, Polianski2001, Avron,Wohlman, Aharony, Cremers, Polianski2002, Moskalets2002, Sharma, Citro, BanerjeeDasRao2003, Moskalets2004, DasRao2004, Sela, Splettstoesser2005, Banerjee2006,
Agarwal,DasShpitalnik2008}. There has also been some work\cite{Wang2001, Blaauboer2002, Wang2002-65, Wang2002-66, Taddei, Governale,Kopnin, Splettstoesser2007, Russo, SahaDas2008,
Gibertini, Palop, Paul, Hekking} on normal metal-superconductor  interfaces including Majorana mediated charge pumps.

There has also been recent interest in cases when the pumped charge is quantized, and
in particular for topological reasons\cite{Gibertini}, so that it is stable to disorder and could be used for metrological applications. As mentioned above,
this was first studied by Thouless\cite{Thouless} who showed that the quantised adiabatic charge transport was
related to the Chern number of the band, which also counts the number of monopoles or
equivalently gapless points enclosed by the pumping contour. In recent work\cite{Gibertini}, it has been
shown that the presence of a single transmitting channel at the interface between a normal
wire and a superconductor enables quantization of  the pumped charge by tuning the 
system through topological phase transitions so that  isolated topological trivial regions
are surrounded by topological regions. Thus pumping paths can be chosen to make non-contractible
loops in the parameter space which could  leads to quantized charge pumping.

In an earlier paper, we studied the conductance through a Majorana bound state (MBS) embedded in a Aharonov-Bohm ring geometry
and showed that the currents at the two leads tunnel coupled to the  Aharonov-Bohm ring were anti-correlated and the degree of anti-correlation could be tuned by
the Aharonov-Bohm ($\mathcal{AB}$) flux threading the ring.
In this paper, we will explore charge pumping through the MBS, in the same geometry and
study the role of the $\mathcal{AB}$ ring geometry which exhibits non-trivial topology.  Unlike Ref.~\onlinecite{Gibertini} where the pumping of charge required going through topological phase transitions as one traverses along the  pumping contour,  here we will show show that it possible to obtain quantized pumped charge even when the superconductor hosting the MBS does not undergo phase transitions. 

The ring geometry plays a crucial role here since we will show that there is no pumped charge when the two leads are just connected to the MBS (normal-MBS-normal or simple two-lead 
geometry without the ring). This is unlike the earlier study of conductance where the anti-correlation existed even in the  two-lead geometry and the 
$\mathcal{AB}$ geometry was required only to provide a tuning parameter for  tuning the degree of anti-correlation. Here, on the contrary, the 
$\mathcal{AB}$ geometry is crucial to get non-zero pumping. In this geometry, we will study the pumped charge at each of the leads  using a scattering matrix approach, restricting ourselves to the adiabatic regime, and we will  compute the pumped charge using
the analog\cite{Blaauboer2002}  of the Brouwer\cite{Brouwer}  formula for a normal-superconductor junction. Finally, we will  show that a Fourier analysis of the pumped charge as  a function of the flux through the ring, leads to a single frequency domination for the MBS, as opposed to many harmonics for an Andreev bound state (ABS); this  can be thought of as a diagnostic for the MBS.

\section{Simple connected geometry with two leads}

The Hamiltonian for two normal leads which are tunnel coupled to an MBS situated at the end of a one dimensional p-wave superconductor\cite{Kitaev} is given by 
\begin{align}
H = \underset{\alpha}{\sum}H_{\alpha} +  H_{T} \label{ham},
\end{align}
where the form of the lead Hamiltonian for the two ($\alpha=1,2$) leads are given by $H_\alpha = \int_{-\infty}^{\infty} dx \psi_{\alpha}^{\dagger}(x)(-i v_{F} \partial_{x}) \psi_{\alpha}(x)$ and the tunneling Hamiltonian is given by 
\begin{align}
  H_{T} &= i \gamma \underset{\alpha}{\sum} ( u_{\alpha} \psi_{\alpha}(x=0) + u_{\alpha}^{\ast} \psi^{\dagger}_{\alpha}(x=0) )~. 
 \label{tunneling}
\end{align}
Here $\gamma$ represents the Majorana fermion operator and $u_{\alpha}$ represents the amplitude of coupling between the lead $\alpha$ and the MBS which is complex number in general. 
The scattering matrix describing the scattering of electrons and holes  between the leads via the MBS corresponding to the situation described by the above tunnel Hamiltonian is found by applying the Weidenmuller formula \cite{Tripathi}  as
\begin{align}
 S(E) &=  \begin{pmatrix}
   S^{ee}(E) & S^{eh}(E)\\
   S^{he}(E) & S^{hh}(E)\\
   \end{pmatrix},
       \end{align}  
       where 
  \begin{align}
 S^{ee}(E) &=  1_{2} - \frac{2i\pi\nu }{d(E)}\begin{pmatrix}
  |u_{1}|^{2} & u_{1}^{\ast}u_{2} \\
  u_{2}^{\ast}u_{1} & |u_{2}|^{2} \nonumber\\
   \end{pmatrix}, \\ \\
 S^{he}(E) &=   -\frac{2i\pi\nu}{d(E)}\begin{pmatrix}
  u_{1}^{2} & u_{1}u_{2} \\
  u_{1}u_{2} & u_{2}^{2}
   \end{pmatrix} \nonumber\\
       \end{align}       
with  $
 d(E) = E + 2i\pi\nu (|u_{1}|^{2}+|u_{2}|^{2})$. Here the scattering matrix is written in a basis where the first two rows and columns correspond to electrons from lead-$1$ and lead-$2$ respectively,  and the next two rows and columns correspond to holes from lead-$1$ and lead-$2$ respectively, and $\nu$ represents the density of states of the electrons which has been assumed to be same in both leads for simplicity.
 
Using the extension of Brouwer's formula for the case of superconducting junction \cite{Blaauboer2002},  the pumped charge at each lead  is given by
\begin{align}
  Q_{\alpha} = -e \int_{{\cal{A}}} dX_{1}dX_{2} Im[C_{\alpha,\alpha}]  \label{chargeQ}
  \end{align}
where the matrix 
\begin{align}
  C = \frac{1}{\pi}[\frac{d}{dX_{1}}S^{ee}\times \frac{d}{dX_{2}}(S^{ee})^{\dagger} - \frac{d}{dX_{1}}S^{he}\times \frac{d}{dX_{2}}(S^{he})^{\dagger}]~. 
   \label{chargeQ1}
  \end{align}
 Here $X_1$ and $X_2$ represent the pumping parameters which are periodic functions of the  time $t$. They trace out a closed loop in the $X_1$-$X_2$  plane over one time period such that the area enclosed by the loop is finite and is given by ${\cal{A}}$. Also note that the pumped charge in each lead can be decomposed into a particle-like process which depends on $S^{ee}$ alone and a particle-hole conversion process which depends on $S^{he}$ alone. 
  
We can choose the $u_\alpha$'s to be  the pumping parameters, $i.e.$,  we can choose $X_1 = u_1$ and $X_2 = u_2$.
In this case, we find that 
\begin{align}
  Im[C_{11}] = 0 = Im[C_{22}]  
 \end{align}
  -i.e., the integrand itself vanishes and there is no pumped charge.
Note that the $u_\alpha$'s can be  taken to be real since their phase can be  gauged away as long as it is not time dependent. Alternatively,  if we intend to use the phase of the $u_\alpha$'s  as a pumping parameter (i.e.,  make it time dependent), the implementation of such a pumping protocol requires the order parameter of the superconducting hosting the MBS to be varied in time which in turn calls for Josephson junction type setup which is beyond the scope of the present work. 

Now, the vanishing of the pumped charge can be attributed to the particle-hole symmetry of the couplings between the MBS and the leads about zero energy due to the fact that the MBS is pinned to zero energy. To contrast the case of MBS to a more regularly encountered  bound state in the context of normal-superconducting hybrid structures, the Andreev bound state (ABS),  we will show now that the vanishing of the pumped charge is not true in general. When we couple  an ABS to leads, we will see that it leads to a finite pumped charge even when the ABS is tuned to zero energy. To evaluate the pumped charge via an ABS, we start by replacing the tunnel Hamiltonian in Eq.[\ref{tunneling}] by \cite{Tripathi}

\beq
H_{T} = a^{\dagger} \underset{\alpha,k}{\sum} ( t_{\alpha} c_{\alpha k} + v_{\alpha}^{\ast} c^{\dagger}_{\alpha k} ) + h.c. , 
\label{tunnelingABS}
\eeq
where now $a^\dagger$ denotes the creation operator for the ABS, (which, unlike the MBS does not have to be real) and the tunneling amplitudes to the electron
and hole states on the leads are given by $t_\alpha$ and $v_\alpha^*$ respectively. 
Considering $t_{1}$ and $t_{2}$  to be 
pumping parameters, and parametrising the contour in terms of a scale $R_2$, 
($t_1^2 + t_2^2 / R_2^{2}  = 1$), we can obtain the pumped charge.

\begin{figure}[t]
\centering
\includegraphics[scale=0.2,width=0.35\textwidth]{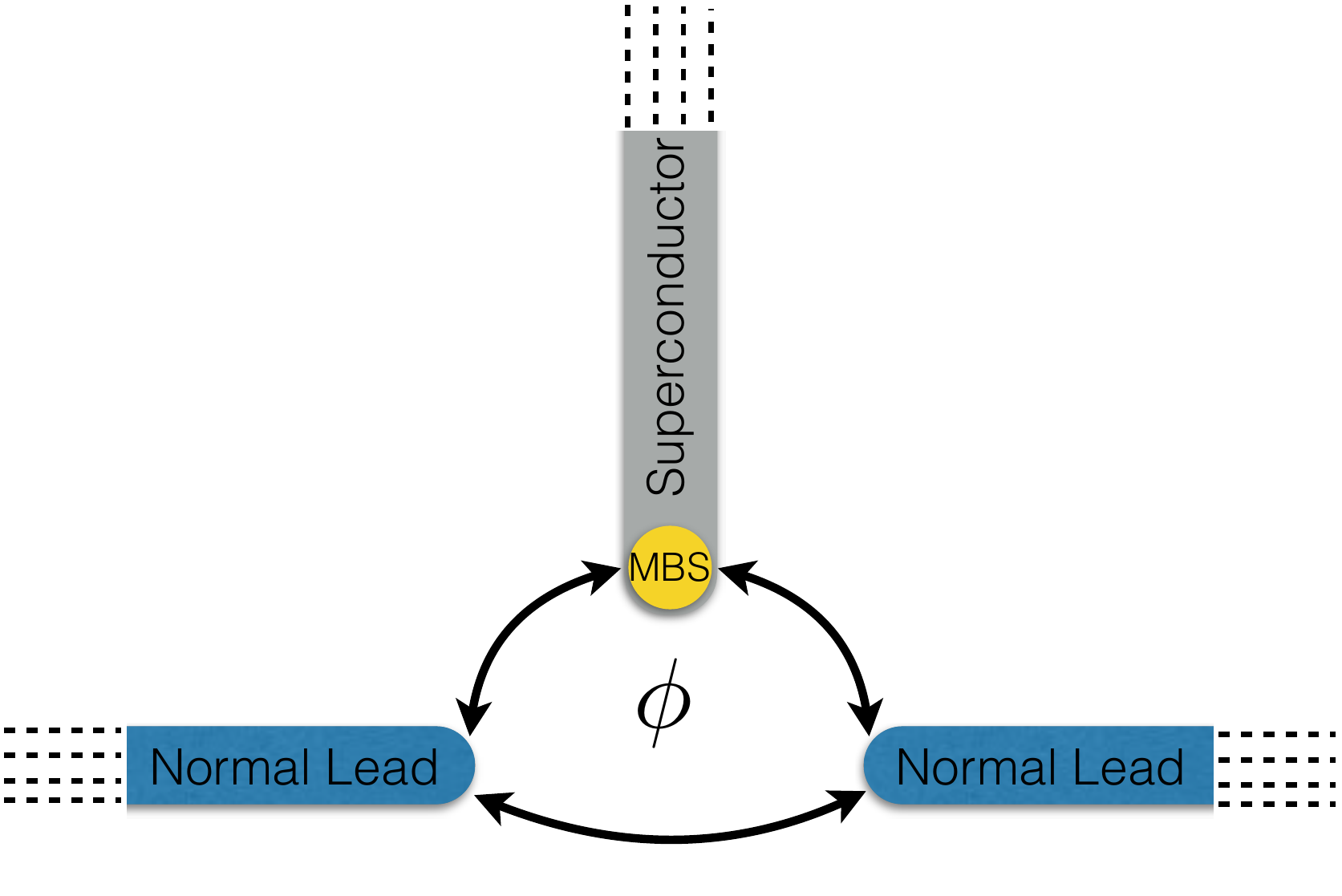}
\caption{ (color online) Schematic illustration  of the $\mathcal{AB}$ setup with two normal leads, which are simultaneously tunnel coupled  to the MBS and to each other. Here the tunnel coupling is represented as a double headed arrow and the flux through the ring type geometry is represented by $\phi$.}
\label{fig1}
\end{figure}

\begin{figure}[htb]
\centering
\includegraphics[width=0.35\textwidth]{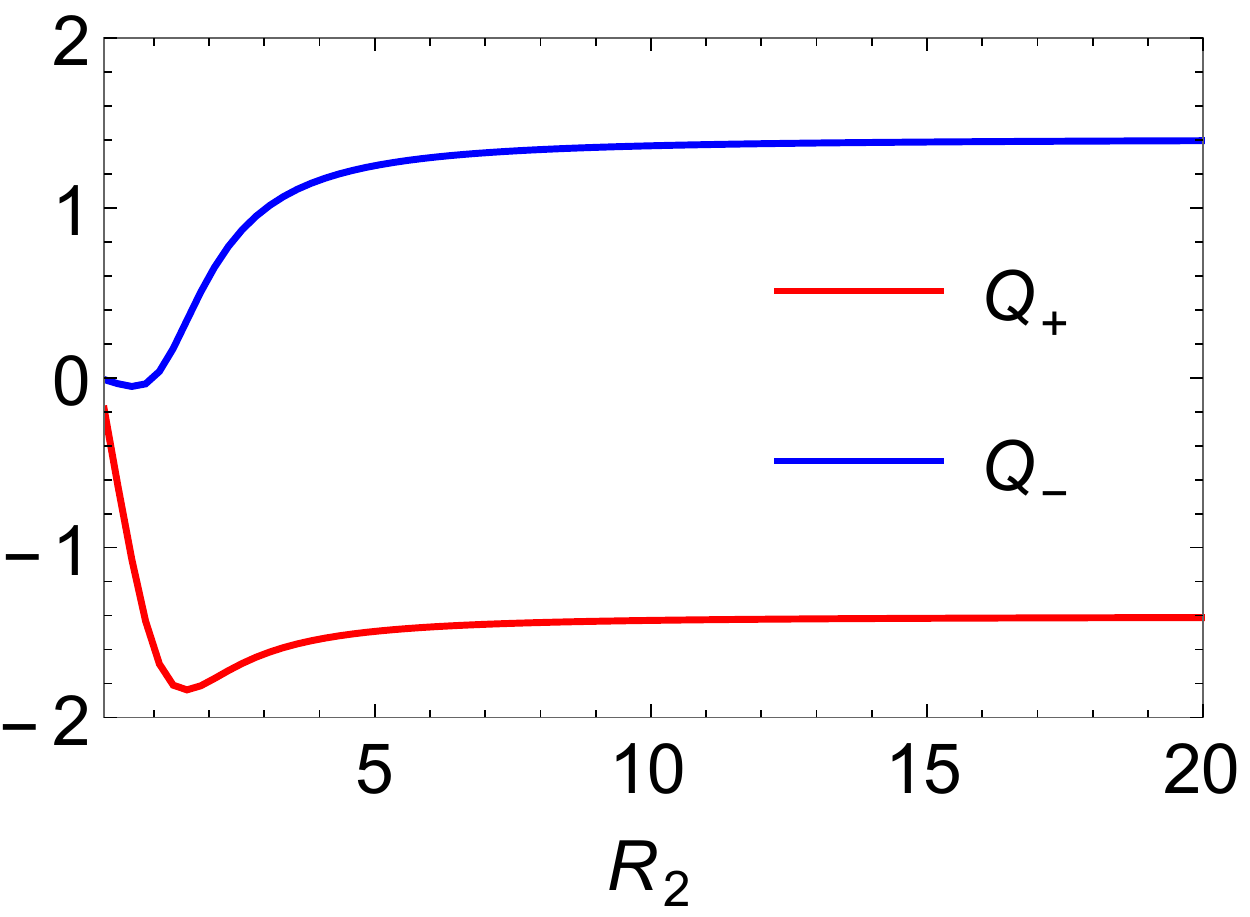}
\caption{ (color online) The charge pumped to the  superconductor, given by $Q_+$, is shown in blue while the charge pumped between the two normal leads through the ABS, given by $Q_-$, is shown in red  (in the absence of the direct tunneling 
between the leads), in units of the electronic charge $e$.  The pumping occurs in the  $t_1 - t_2$ plane and the pumped charge is shown as a  function of a  scale  $R_2$ which parametrises the pumping contour as  $t_1^2 + t_2^2 / R_2^{2}  = 1$ . The parameters  $v_1,v_2$ are chosen to be $v_1=1,v_2=i$}
\label{fig2}
\end{figure}

As the analytic expression for the pumped charge in this case gets cumbersome,   we perform a numerical analysis
for some representative values  to obtain it  using Eq.[\ref{chargeQ},\ref{chargeQ1}]. This is presented in Fig.\ref{fig2}. Note that the total charge pumped from the normal metal leads into the superconductor is given by $Q_{+}=Q_1+Q_2$ and the total charge pumped from lead-$1$ to lead-$2$ via the MBS is given by $Q_{-}=Q_1-Q_2$ over a single pumping cycle. We observe that both these quantities are finite for the chosen pumping contour and they asymptotically reach a steady value as the amplitude of pumping parameter($R$) gets larger and larger. Hence, this fact itself,  presents a clear distinction between the ABS and MBS. 

\section{Multiply connected geometry and the MBS}

From the above analysis it is clear that a simple set up involving two leads tunnel coupled to an MBS does not lead to net charge being pumped either from one lead 
to another or from the leads to the superconductor over a pumping cycle. Next, we explore the possibility of pumping in a multiply connected geometry where the MBS is embedded in a  ring. 

\vspace{0.2cm}

\noindent{{\it MBS  embedded in an $\mathcal{AB}$ geometry:-}} We now look for a multiply connected geometry and the simplest choice is to consider a ring geometry where a magnetic flux is piercing the ring. The ring geometry is realized by considering an MBS which is tunnel coupled to two leads that are also directly tunnel coupled to one another as shown in Fig.\ref{fig1}. The Hamiltonian for the system is the same as that given in Eq.[\ref{ham}]
except that we now have an additional direct tunneling term given by 
\begin{align}
  H_{direct} =  \tau \psi^{\dagger}_{1}(x=0)\psi_{2}(x=0) + h.c.)~. \label{Majtunneling}
\end{align}
Here $\tau$ denotes the amplitude for the direct tunnel coupling of the two leads to each other.  The scattering matrix now involves
the direct coupling term as well and is given by
 \begin{align}
 S^{ee}(E) &=  \frac{1}{1+\pi^{2}\nu^{2}|\tau|^{2}}\begin{pmatrix}
  1-\pi^{2}\nu^{2}|\tau|^{2} & -2i\pi\nu \tau\\
   -2i\pi\nu \tau^{\ast} & 1-\pi^{2}\nu^{2}|\tau|^{2}\\
   \end{pmatrix} \nonumber\\
      &- \frac{2i\pi\nu }{(1+\pi^{2}\nu^{2}|\tau|^{2})^{2}D(E)}\begin{pmatrix}
  u_{1+}^{\ast}u_{1-} & u_{1+}^{\ast}u_{2-} \\
  u_{2+}^{\ast}u_{1-} & u_{2+}^{\ast}u_{2-}\\
   \end{pmatrix} \nonumber\\
       \end{align}  
  and
  \begin{align}
 S^{he}(E) &=   -\frac{2i\pi\nu }{(1+\pi^{2}\nu^{2}|\tau|^{2})^{2}D(E)}\begin{pmatrix}
  u_{1+}u_{1-} & u_{1+}u_{2-} \\
  u_{2+}u_{1-} & u_{2+}u_{2-}\\
   \end{pmatrix} \nonumber \label{DE} \\
       \end{align}     
with  $D(E) = E + \frac{2i\pi\nu }{(1+\pi^{2}\nu^{2}|\tau|^{2})}{(|u_{1}|^{2}+|u_{2}|^{2})}$, 
   $ u_{1s}  = u_{1} + s i \pi\nu \tau^{\ast} u_{2} $, $
   u_{2s}  = u_{2} + s i \pi\nu \tau u_{1}$ and $s=+, -$. 
Once again, the pumped charge can be computed using the analog of Brouwer's formula. Here, we find that
the integrand is given by 
\begin{align}
  Im[C_{11}] &= \frac{16 \pi^{2}\nu^{3}\tau_{0} (1+ \pi^{2}\nu^{2}\tau_{0}^{2}) \cos(\phi) E^{2} (u^{2}_{1} + u^{2}_{2})}
                {[E^{2} (1+ \pi^{2}\nu^{2}\tau_{0}^{2})^{2} + 4 \pi^{2}\nu^{2} (u^{2}_{1} + u^{2}_{2})^{2}]^{2}}  \nonumber \\
             &=  -Im[C_{22}]~. 
 \end{align}
As before, the  $u_\alpha$s are taken to be real,  and the direct tunneling term is taken to be  $\tau = \tau_{0} e^{i \phi}$ where $\phi$ plays the role of the $\mathcal{AB}$ flux.
Clearly, this expression is zero when the direct tunneling amplitude is zero and it agrees with the earlier result.
Using this in Eq.[\ref{chargeQ}], we see that the integrand and consequently, the pumped charge through each lead is zero 
even in the presence of a direct tunneling term, at $E=0$. 

So although, we have allowed for a  finite direct tunneling amplitude ($\tau_0$) for the electrons, leading to a multiply connected geometry,  we are still unable to break the particle-hole symmetry of the pumped charge about $E=0$, which prevents pumping of net charge. The MBS in many aspects is very similar to a resonant level (RL). The MBS allows for resonant injection of a pair of electron into superconductor via resonant Andreev process and the RL allows for a single electron to resonantly transmit across it. Hence to gain insight in the MBS pumping analysis, our next analysis is to consider pumping of charge across a RL embedded into an $\mathcal{AB}$ geometry.

\vspace{0.2cm}
\noindent{{\it $\mathcal{AB}$ geometry and the resonant level:-}} We can now contrast the above observed behaviour of the MBS to the  case where  the MBS is replaced by a  RL. The only change in the above model is that  the first line of the tunneling term in Eq.[\ref{Majtunneling}]  that represents tunneling through the MBS is now replaced by 
 \begin{align}
  H_{T} &= (d^{\dagger} \underset{\alpha}{\sum} u_{\alpha} \psi_{\alpha}(x=0) + h.c.) 
\end{align}
where $d^\dagger$ represents the creation operator of the electron on the resonant level. The direct coupling term between the leads remains the same.
 The scattering matrix in this case is given by:
 \begin{align}
 S(E) &=  \begin{pmatrix}
   S_{11}(E) & S_{12}(E)\\
   S_{21}(E) & S_{22}(E)\\
   \end{pmatrix} \nonumber \\
S_{ij;i=j}(E) &= -1 + \frac{2(E + i \pi\nu u_{ij})}{{\tilde d}(E)}       \nonumber \\
S_{ij;i \neq j}(E) &=  -\frac{2 i \pi\nu(E \tau_{0} e^{i\phi{ij}} +  u_{1}u_{2})}{{\tilde d}(E)}       \nonumber \\
{\tilde d}(E) &= E(1+ \pi^{2}\nu^{2}\tau_{0}^{2}) + 2 \pi^{2}\nu^{2} \tau_{0}\cos (\phi) u_{1}u_{2} \nonumber \\
     &+  i \pi\nu (u^{2}_{1} + u^{2}_{2}) ,   \label{dE}
\end{align}
where $u_{11}=u^{2}_{2}, ~u_{22}=u^{2}_{1}$ and $\phi_{12}=\phi$,  $\phi_{21}=-\phi$.
Now the pumped charge can be evaluated from the expressions for $Im[C_{11}]$ and $Im[C_{22}]$ as given below,
  
\begin{align} \label{Integrand-RL}
 Im[C_{11}] &= \frac{8 \pi^{2}\nu^{3}}{|d(E)|^{4}} (u^{2}_{1} + u^{2}_{2}) [E^{2} \tau_{0} \cos(\phi) (1+ \pi^{2}\nu^{2}\tau_{0}^{2}) \nonumber \\
            & ~~~+ E \{ u_{1}u_{2} (1+ \pi^{2}\nu^{2}\tau_{0}^{2} \cos(2\phi)) \nonumber \\
            &~~~ -  \pi\nu\tau_{0} \sin(\phi) (u^{2}_{1} - u^{2}_{2}) \}]  \nonumber \\
            &=  -Im[C_{22}]~.
\end{align} 
Note that unlike the MBS case, where the integrand vanishes without direct tunneling between the leads, here the integrand is non-zero even for $\tau_0 =0$. This clearly indicates that although  the direct tunneling term or the ring geometry was absolutely necessary  to even get a non-zero integrand for the MBS case, that is not the case  for the RL case. However, the pumped charge through each lead continues to be identically zero in both cases at $E=0$.
\vspace{0.2cm}

\begin{figure}[htb]
\centering
\begin{subfigure}
\centering
\includegraphics[width=0.235\textwidth]{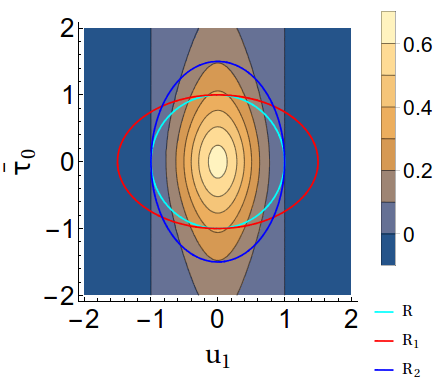}
\end{subfigure}
\begin{subfigure}
\centering
\includegraphics[width=0.235\textwidth]{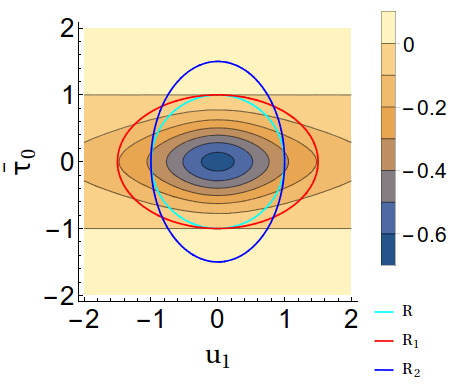}
\end{subfigure}
\caption{ (color online) The integrands $Im(C_{11}+C_{22})$ and $Im(C_{11}-C_{22})$ are plotted as a function of the
pumping parameters $u_1$ and $\bar\tau_{0}$. Here, we have taken $u_2=1$ and the phase of the direct hopping  $\phi=0$. The three 
contours $a$,$b$ and $c$ (explained in the text) for which  we have computed the pumped charges $Q_{+}$ and
$Q_{-}$  in Fig.(\ref{fig4}) are shown in cyan, blue  and red respectively.}
\label{fig3}
\end{figure}

\noindent{{\it $\mathcal{AB}$ geometry and the role of pumping parameters -}} We note above that for both the MBS and the RL,  the pumped charge is zero at $E=0$,  while it is finite in general for ABS. Hence it is natural is ask whether the choice of pumping parameters can change this fact. For  the case of the MBS embedded in a ring we replace the pumping parameter $u_{2}$, which is one of the hopping amplitudes to the MBS, by $\bar\tau_{0} = \pi \nu \tau_{0}$ which is the amplitude for direct tunneling between the leads.  Note that this explicitly breaks the symmetry between the two leads as far as the pumping contour is concerned. In this case, we find that even at $E=0$ the integrand  is finite and is given by 
\begin{align}  \label{Integrand-MBS} 
  Im[C_{11}] &= \frac{2 \cos(\phi) u_{2} [u_{1}^{2} - \bar\tau_{0}^{2} u_{2}^{2}]} 
                {\pi(u_{1}^{2} + u_{2}^{2})^{2}(1 + \bar\tau_{0}^{2})^{2}}  \nonumber \\
  Im[C_{22}] &=  -\frac{2 \cos(\phi) u_{2} [u_{2}^{2} - \bar\tau_{0}^{2} u_{1}^{2}]}
                {\pi(u_{1}^{2} + u_{2}^{2})^{2}(1 + \bar\tau_{0}^{2})^{2}} ~. 
 \end{align}
 In Fig.\ref{fig3}, we show a plot of the integrands $C_{11}$ and $C_{22}$ as a function of
 the two pumping parameters. The pumped charge at the two leads can now be computed
 by choosing various contours. We show below the pumped charge for a few representative
 contours and note how asymptotically, (almost) quantised charge is pumped either to 
 the superconductor or between the two leads.

\vspace{0.2cm}

\noindent{{\it Pumped charge for various pumping contours:-}}
As the physically relevant quantities are $Q_{+}$ and $Q_{-}$ , we first plot the integrands $Im(C_{11}+C_{22})$ and $Im(C_{11}-C_{22})$ as a function of the
pumping parameters $u_1$ and $\bar\tau_{0}$ as shown in Fig.\ref{fig3}. Note that the maximum value for $Q_{+}$ and $Q_{-}$ are concentrated about the $u_1=0$ and  $\bar\tau_{0}=0$ axis respectively. Also, note that the sign of $Im(C_{11}+C_{22})$ and $Im(C_{11}-C_{22})$ remains the same as we move along the axis about which the maxima of  these functions are mostly distributed. On the other hand  $Im(C_{11}+C_{22})$ and $Im(C_{11}-C_{22})$ do change sign along the axis perpendicular to the axis of the distribution of the  maxima. This fact will strongly influence the asymptotic values of the pumped charge as we go to larger and larger contour sizes. For obtaining large values of pumped charge we need to analyze the symmetries of the distribution of 
values of $Im(C_{11}+C_{22})$ and $Im(C_{11}-C_{22})$ and design  pumping contours which will efficiently enclose
a large fraction of the  maxima of these functions in the parameter space. In principle, appropriately chosen contours can lead to asymptotically quantized value for pumped charge\cite{Levinson, BanerjeeDasRao2003, DasRao2004, DasShpitalnik2008,SahaDas2008}.  Keeping this fact in mind we consider elliptical shapes of the contours in the plane of pumping parameters $(u_1,\bar\tau_{0})$  given by $ u_1^2/R_1^{2} +\bar\tau_{0}^2/R_2^{2} = 1$.  

\begin{figure}[htb]
 \centering
 \includegraphics[width=0.35\textwidth]{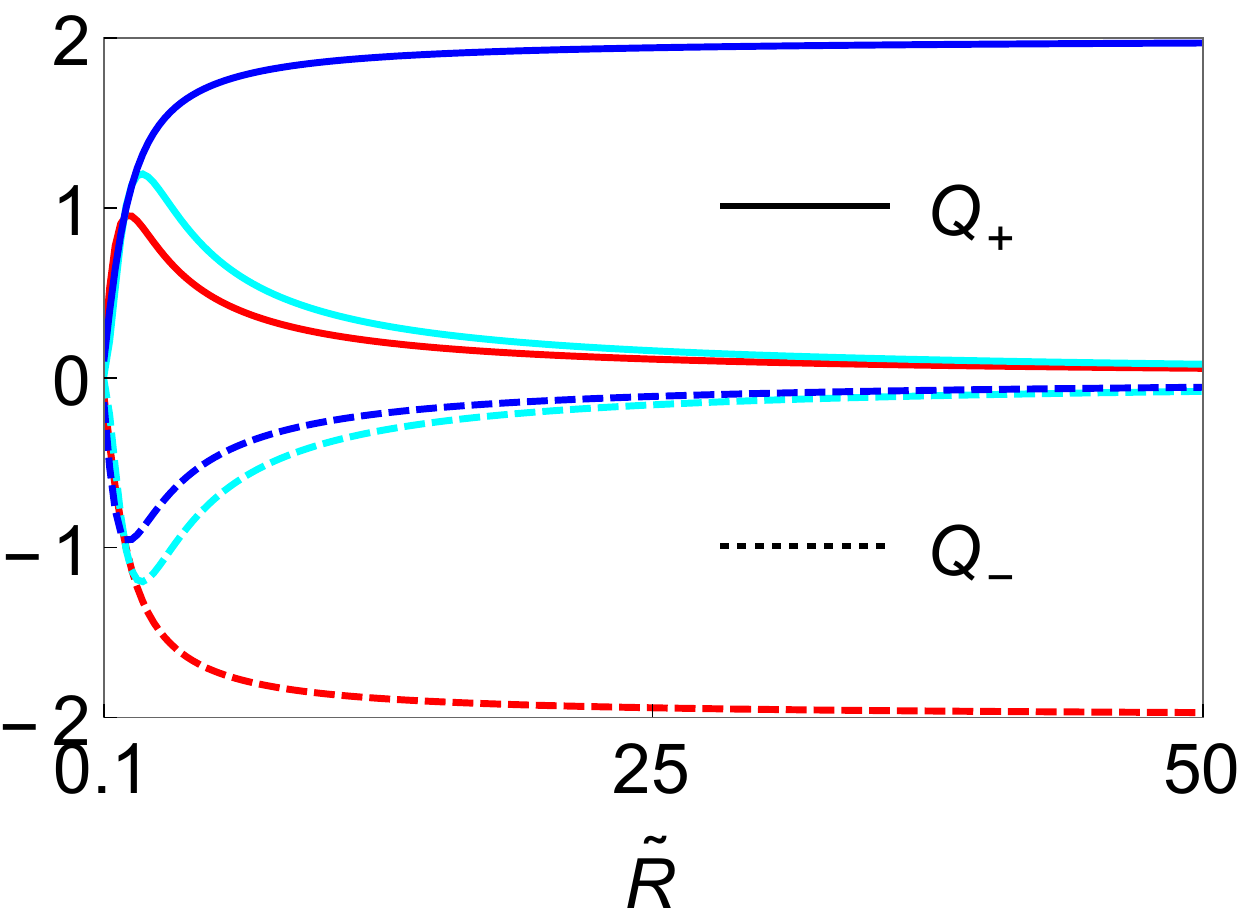}
 \caption{ (color online) Pumped charges to the superconductor and between the leads($Q_+$ and $Q_-$) for the MBS in the presence of  direct tunneling in units of electronic charge $e$ for pumping in $u_1 - \bar\tau_{0}$ plane as a 
function of the  scale of the relative pumping strengths (the elliptic contours) and the radius for equal pumping strengths  (circular contour). More explicitly, the variable $\tilde R$ which labels the $x$-axis denotes
$R, R_1$ and $R_2$ respectively for the contours $(a)$,$(b)$ and $(c)$ respectively. The parameter $u_2$ is chosen to be   $u_2=1$.}
 \label{fig4}
 \end{figure}

We have produced plots for three different kinds of contours which are given by $(a)$ $R_1=R_2= R$ where the  asymptotic pumped charge is obtained for $R\rightarrow \infty$ limit,  $(b)$ $R_1/R_2 >1$ where the asymptotic pumped charge is obtained for $R_1\rightarrow \infty$ limit, and $(c)$  $R_2/R_1 >1$ where the  asymptotic pumped charge is obtained for $R_2\rightarrow \infty$ limit. In Fig.\ref{fig3} we have shown representative contours for the cases $(a)$, $(b)$ and $(c)$ discussed above in  cyan, blue and red respectively. The corresponding asymptotic pumped charge is given in Fig.\ref{fig4} where the color code of corresponding cases are kept the same. 

Let us first discuss the results corresponding to the $(b)$-type contour which is depicted in blue in Figs.\ref{fig3} and \ref{fig4}. We note that  $Q_{+}\rightarrow 2$, i.e., gets asymptotically quantized while $Q_{-}\rightarrow 0$ as $R_1\rightarrow \infty$.  This fact is consistent with our observation that the maximum of $Im(C_{11}+C_{22})$ is distributed around the $u_1=0$ axis and the $(b)$-type contour maximally encloses the area around this axis, hence leading to quantization of $Q_{+}$. On the other hand, $Im(C_{11}-C_{22})$ changes sign as we move along the $u_1=0$ axis and hence $Q_{-}$ shows a non-monotonic behaviour and finally goes to zero as $R_1\rightarrow \infty$. 

The same logic can be used to understand the fact that $(a)$-type contour always shows a non-monotonic behaviour for the pumped charge which always goes to zero in the asymptotic limit. This is so because the $(a)$-type contour always engulfs areas where $Im(C_{11}+C_{22})$ and $Im(C_{11}-C_{22})$  both undergo sign changes,  hence cancelling to zero in the  asymptotic limit. Finally it is clear from the above arguments that  the $(c)$-type contour will show a behaviour which is exactly complementary to the $(b)$-type contour since  the maximum values of $Im(C_{11}+C_{22})$ and $Im(C_{11}-C_{22})$ are distributed around complementary axis ( i.e., $u_1=0$ and $\tau=0$ axis respectively). 
Hence we have shown that by choosing appropriate contours we are in a position to selectively pump charge from the leads to the superconductor ($Q_{+} \neq 0$) while keeping the relative transfer of charge between the leads to be zero ($Q_{-}=0$) or pump charge between the leads while keeping the superconductor decoupled (i.e., $Q_{+} =0$).

\vspace{0.2cm}
 
\begin{figure}[htb]
\centering
\includegraphics[width=0.37\textwidth]{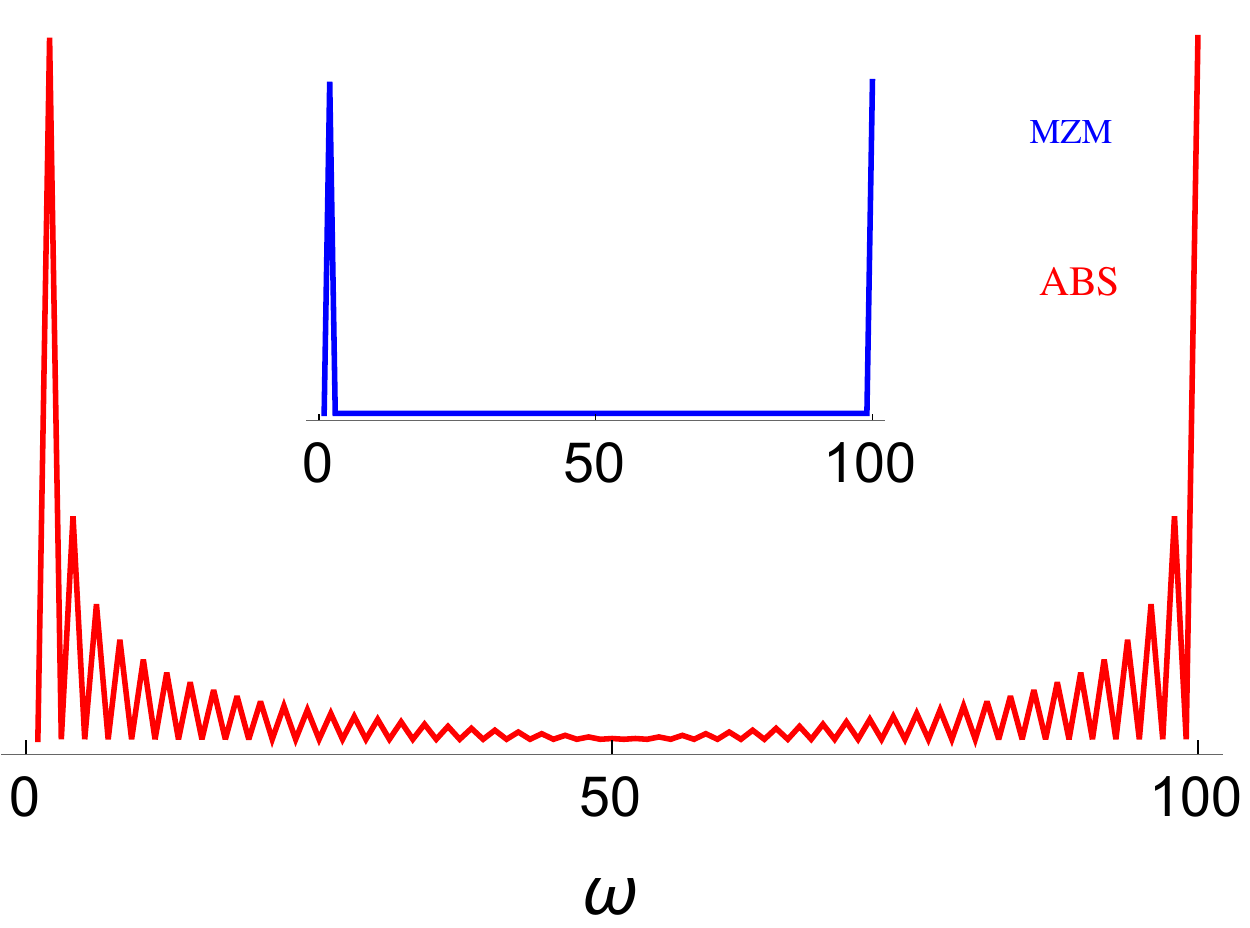}
\caption{ (color online) Plot of discrete Fourier transform of $Q_+$, $|A(\omega)|^{2} = |\frac{1}{N} \sum_{\phi} Q[\phi] e^{i 2*\pi* \omega \phi/N}|^2$ as a function of the  frequency $\omega$ for pumping in $u_1 - \bar\tau_{0}$ plane for ABS in red and MBS in blue.  The contour is chosen to be circular with $R=25$. The number of points $N$ is chosen to be 100 and the other parameters are given by
 $u_{2}=1$, $v_{1}=3$, $v_{2}=2$.
}
\label{fig5}
\end{figure}

\noindent {\it Fourier analysis of pumped charge:-} Finally we would like to point out a crucial difference in the scattering matrix for the MBS in the multiply connected geometry and other forms of bound states like the ABS or the RL in the same geometry. The other form of bound states in general would lead to a $\phi$ dependent denominator which appears due to  the Fabry-Perot type interference due to the circulating paths of electrons or holes around the multiply connected geometry. But due to the fine tuned symmetry between an electron and a hole for the MBS, all such phases cancel out to provide a $\phi$ independent denominator. This can be seen clearly by comparing the expression for $D(E)$ in Eq.\ref{DE} with ${\tilde d}(E)$ in Eq.\ref{dE} . Also note that the $\phi$ dependence for MBS appears in the numerator of the scattering matrix as a pure cosine. The same difference in dependence also persists
in the expression for the integrand of the pumped charge, as can be seen from Eqs.\ref{Integrand-RL} and \ref{Integrand-MBS}. This essentially means that the pumped charge for 
the MBS has a single 
periodicity with respect to variation of $\phi$ as opposed to the RL or the ABS which will have superperiods in $\phi$.  This can serve as a diagnostic for the MBS. This fact can be seen very clearly from a Fourier analysis for the pumped charge shown in Fig.\ref{fig5}. 

As expected the MBS case show a clear delta function like peak which is independent of the parameters chosen for the analysis due to the fact that only the first of the harmonics contributes to this case, whereas for the ABS, there are multiple frequencies signifying the presence of higher harmonics which can be traced back to the $\phi$ dependent expression for $d(E)$.

\vspace{0.2cm}

\noindent {\it Discussions and conclusion :}

In this letter, we have discussed the importance of a ring geometry to get non-zero pumped charge through the MBS. We note that asymptotically two units of charge can either be pumped between the leads or from the leads to the superconductor. We do not get quantised single unit charge pumping within
our setup, because our pumping protocol only involves the Majorana bound state at one end of the topological superconductor.  Hence, the fermion parity is fixed.

We then show that the Fourier analysis of the pumped charge  through  an $\mathcal{AB}$ ring,
can be used as a diagnostic to distinguish between MBS from other spurious zero energy states. In particular, the charge pumped through the   MBS is different from the charge pumped through either
 the resonant level or the  ABS in that it has no higher harmonics. 
 This is true   independent of choice of the contour, and is consequently  a strong diagnostic.

\noindent {\it Acknowledgments} 

The research of K.M.T was supported in part by the INFOSYS scholarship for senior students.

\renewcommand{\thefigure}{A\arabic{figure}}
\setcounter{figure}{0}
\renewcommand{\theequation}{A\arabic{equation}}
\setcounter{equation}{0}

\end{document}